\documentclass{PoS}

\usepackage[varg]{txfonts}
\usepackage{multirow}
\usepackage{url}
\usepackage{soul}

\usepackage{graphicx}
\usepackage{subfig}

\newcommand{\Swift}{\textit{Swift}}
\newcommand{\RXTE}{\textit{RXTE}}

\title{Spectral characteristics of Mrk\,501 during the 2012 and 2014 flaring states}

\ShortTitle{Spectral characteristics of Mrk\,501}

\author{\speaker{Gabriele Cologna}\,$^a$\,\thanks{Member of the International Max
    Planck Research School for Astronomy and Cosmic Physics at the University
    of Heidelberg (IMPRS-HD) and the Heidelberg Graduate School of Fundamental
    Physics (HGSFP).} , Nachiketa Chakraborty\,$^b$, Mahmoud
  Mohamed\,$^a$\,\footnotemark[2] , Frank Rieger\,$^b$, Carlo Romoli\,$^c$,
  Andrew Taylor\,$^c$, Stefan J. Wagner\,$^a$, Alicja
  Wierzcholska\,$^{a,\,d}$\,\thanks{Mobility Plus Fellow} , Agnieszka
  Jacholkowska\,$^e$ for the H.E.S.S. Collaboration and Omar
  Kurtanidze\,$^{f,\,a}$\\ 
        \llap{$^a$} Landessternwarte, Universit\"at Heidelberg, K\"onigstuhl, D 69117 Heidelberg, Germany \\
        \llap{$^b$} Max-Planck-Institut f\"ur Kernphysik, P.O. Box 103980, D 69029 Heidelberg, Germany \\
        \llap{$^c$} Dublin Institute for Advanced Studies, 31 Fitzwilliam Place, Dublin 2, Ireland \\
        \llap{$^d$} Insitute of Nuclear Physics, Polish Academy of Science, ul. Radzikowskiego 152, 31$-$342 Krak\'{o}w, Poland\\
        \llap{$^e$} LPNHE, Universit\'e Pierre et Marie Curie Paris 6, Universit\'e Denis Diderot Paris 7, CNRS/IN2P3, 4 Place Jussieu, F-75252, Paris Cedex 5, France\\
        \llap{$^f$} Abastumani Observatory, Mt. Kanobili, 0301 Abastumani, Georgia\\
        E-mail: \email{gcologna@lsw.uni-heidelberg.de}}

\abstract{Observations at Very High Energies (VHE, E\,>\,100\,GeV) of the
  BL\,Lac object Mrk\,501 taken with the High Energy Stereoscopic System
  (H.E.S.S.) in four distinct periods between 2004 and 2014 are
  presented, with focus on the 2012 and 2014 flaring states. The
  source is detected with high significance above $\sim$2\,TeV in
  $\sim$13.1\,h livetime. The observations comprise low flux states and strong
  flaring events, which in 2014 show a flux level comparable to the 1997
  historical maximum. Such high flux states enable spectral variability and
  flux variability studies down to a timescale of four minutes in the
  2-20\,TeV energy range. During the 2014 flare, the source is clearly
  detected in each of these bins. The intrinsic spectrum is well
    described by a power law of index $\Gamma=2.15\pm0.06$ and does not show
    curvature in this energy range. Flux dependent spectral analyses show a
  clear harder-when-brighter behaviour. The high flux levels and the
    high sensitivity of H.E.S.S. allow studies in the unprecedented
  combination of short timescales and an energy coverage that extends
  significantly above 10\,TeV. The high energies allow us to probe the effect
  of EBL absorption at low redshifts, jet physics and LIV. The multiwavelength
  context of these VHE observations is presented as well.} 

\FullConference{The 34th International Cosmic Ray Conference,\\
		30 July- 6 August, 2015\\
		The Hague, The Netherlands}

\begin{document}

\section{Introduction}

With the exception of two starburst galaxies and a few sources in the Large
Magellanic Cloud, active galactic nuclei (AGN) are up to now the sole detected
extragalactic TeV emitters. Despite several classes of AGN existing, the
majority of the detections in this energy range consists of blazars, in
particular the subclass of high frequency peaked BL\,Lac objects
(HBL). Mrk\,501 is one of the nearest HBL, with a redshift $z\sim$\,0.034, and
was the second extragalactic object to be detected as Very High Energy (VHE,
E$>$100\,GeV) emitter in 1995 \cite{Whipple1995_1996ApJ...456L..83Q}. It is
known for its strong variability at all energies and is referred to as an
extreme HBL since its synchrotron peak shifts to up to two orders of magnitude
higher energies during flares \cite{Pian1997_1998ApJ...492L..17P}. Thus, a
flux dependent spectral hardening has been observed in both X-rays and in the GeV-TeV band
(e.g. \cite{Pian1997_1998ApJ...492L..17P,CAT_1997_1999A&A...350...17D,FERMI_mwl2008_2009_2011ApJ...727..129A}). 
Because of these characteristics, Mrk\,501 has been observed extensively at
VHE, often during multiwavelength (MWL) campaigns to try to characterize its
SED with simultaneous data, which are of primary importance for such a rapidly
variable source. In 1997, it was found in a high state with fluxes
$\gtrsim$1\,Crab Unit (c.u.) for several months
(e.g. \cite{Whipple1997_1997ApJ...487L.143C,Whipple1997_1998ApJ...501L..17S,HEGRA_1997A&A...327L...5A,HEGRA1997_spectrum_1999A&A...349...11A,HEGRA1997_temp_char_1999A&A...342...69A,CAT_1997_1999A&A...350...17D}) 
%,Milagro1997_1999ApJ...525L..25A,TelescopeArray1997_1998ApJ...504L..71H,TIBET_1997_2000ApJ...532..302A})
and the historically highest VHE flux
(F(>\,250\,GeV)\,$\sim$\,8.3$\cdot$10$^{-10}$\,cm$^{-2}$\,s$^{-1}$) was
recorded on April 16, 1997 \cite{CAT_1997_1999A&A...350...17D}. Such a
prolonged high state has not been registered since then, but important flaring
activities have been observed on a few occasions, like in 2005
\cite{MAGIC_2005_2007ApJ...669..862A}, 2012 \cite{MAGIC_ICRC_2013} and 2014
(this work).

\section{Observations and Analysis}

\textbf{H.E.S.S.} The H.E.S.S. observations have been carried out in four
different epoches between 2004 and 2014. The dataset comprises 34
runs\footnote{The term "run" refers to a single observation, with a typical
  exposure of 28 minutes.} taken in wobble mode with the HESS\,I array in 2004
(4), 2006 (5) and 2012 (4) and with the full HESS\,II array in 2014 (21). The
2004 and 2006 runs were part of MWL campaigns
%\cite{Gliozzi_2006ApJ...646...61G,MAGIC_2006_2009ApJ...705.1624A} 
and were taken on the nights of June 15-16, 2004 and July 18-19, 2006. Due to
the low state of the source, only upper limits could be derived in the
original analysis and the results were published separately
\cite{Aharonian_1st_ul_2005A&A...441..465A,Aharonian_2nd_ul_2008A&A...478..387A}. 
The 2012 and 2014 observations were triggered
%\footnote{Private communication, within the agreement between H.E.S.S., MAGIC
%(MAGIC/FACT in 2014) and VERITAS for notifications about ongoing VHE flaring
%activities of AGN.}
by flaring states in the TeV domain detected by MAGIC and FACT, respectively.
In 2012 two runs in two consecutive nights were taken (June 11 and 12), while
in 2014 Mrk\,501 was observed on several nights between June 19-25 and on July 29-30.

In order to have an homogeneous dataset, only events detected with the CT1-4
telescopes were extracted from the 2014 observations, discarding the
information coming from CT5.
All but four runs pass the standard quality
cuts, for a total livetime of 13.1\,h (12.1\,h when corrected for
acceptance). The mean zenith angle is 63.7$^\circ$ and the mean offset from
the pointing position 0.50$^\circ$. Data reduction has been performed using
the Model analysis \cite{deNaurois2009APh....32..231D} with \textit{Loose}
cuts, chosen to lower the energy threshold as much as possible. The
\textit{Reflected Region Background} method \cite{Berge2007A&A...466.1219B}
was used for the background determination of the spectral analysis. An
  excess of more than 2600 photons is detected with high significance. Thanks
to the higher sensitivity of the Model analysis, Mrk\,501 is detected also in
the years 2004 and 2006. Spectral analyses have been carried out on different
data subsets using the forward-folding technique
\cite{Piron_2001A&A...374..895P} and three different spectral shapes.

\textbf{\RXTE\, and \Swift-XRT.} X-ray coverage has been ensured by \RXTE\,
and \Swift-XRT. The lightcurve of the former has been retrieved from
\cite{RXTE_2013ApJ...772..114R},
%\footnote{\url{http://cass.ucsd.edu/~rxteagn/}},
while the data of the latter collected in WT mode between May 12 and July 30,
2012 (obsIDs: 00030793177-00030793201) and between May 31 and June 28, 2014
(obsIDs: 00035023035-00035023047) have been analysed for this work.
The HEASoft software package
v.\,6.16\,\footnote{\url{http://heasarc.gsfc.nasa.gov/docs/software/lheasoft}}
with CALDB v.\,20140120 has been used. All the events were cleaned and
calibrated using the \verb|xrtpipeline| task and the data in the energy range
0.3-10\,keV with grades 0-2 were analysed. 
The flux values for the lightcurve (Fig.\,\ref{fig:lightcurves}) were
calculated from the spectra of single snapshots integrating between 2 and
10\,keV. These were derived in the following way: the data were grouped using
the \verb|grappha| tool to have a minimum of 30 counts/bin and then fit using
XSPEC v.\,12.8.2 with a single power-law model and Galactic hydrogen
absorption fixed to $n_H=1.58\cdot 10^{20}$\,cm$^{-2}$ \cite{Kalberla05_gal_abs}.
A clear harder-when-brighter behaviour is detected, as also expected from the 
literature (e.g. \cite{Pian1997_1998ApJ...492L..17P}).
Contemporaneous X-rays and H.E.S.S. observations exist for some
nights, and are typically offset by $\sim$\,90\,minutes. 
On June 12, 2012, \Swift-XRT observations start immediatly after the H.E.S.S. ones have finished.

\textbf{Abastunami Observatory.} Optical data are obtained with the 70\,cm
telescope of the Abastunami Observatory (Georgia), equipped with an Apogee 6E
camera (between 1997-2006 a SBIG ST-6 camera). Observations taken with a R
Cousins filter cover the 1997-2014 period and have been analyzed with the
Daophot II reduction software using an aperture diameter of 10$^{''}$. Some
nights of contemporaneous data with H.E.S.S. exist, and strictly simultanous
observations have been taken during the flare peak on June 23, 2014.

\section{Results and discussion}
\label{sec:results}

\begin{figure}[tp]
  \centering
  \includegraphics[width=1.0\columnwidth]{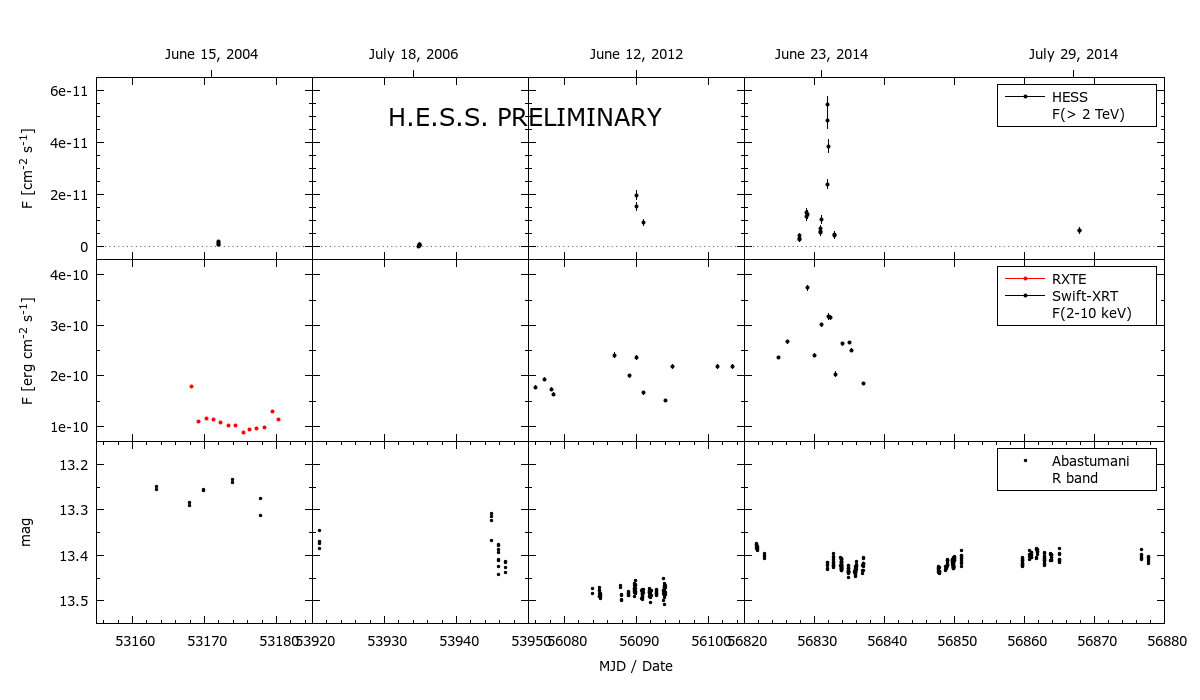}
  \caption{\textit{Top}: H.E.S.S. run-wise lightcurve above 2\,TeV. The flaring
    states in 2012 and 2014 are clearly visible (MJD $\sim$\,56089-90 and
    56831-32). It is also apparent that the "low" state in 2014 still
    represent a state of moderate activity when compared to the flux levels in
    2004 and 2006. \textit{Middle}: \RXTE\, and \Swift-XRT X-ray
    lightcurves in the energy range 2-10\,keV showing high
    variability. \textit{Bottom}: optical lightcurve from the Abastunami
    Observatory. No particularly high flux values or strong variability are
    detected in correspondence of the 2012 and 2014 TeV flares.} 
  \label{fig:lightcurves}
\end{figure}
\textbf{Lightcurves.} In the top panel of Fig.\,\ref{fig:lightcurves}, the
run-wise lightcurve of the 2004-2014 H.E.S.S. observations above 2\,TeV is
shown. Two clear flaring events can be seen in 2012 and in 2014, along with
two quiescent states in 2004 and 2006 and a phase of moderate activity in
2014. Mrk\,501 is clearly variable also in X-rays (middle panel) and in
optical (lower panel). The comparison of the three lightcurves shows no
obvious correlation in the flux variations of the different energy bands. The
optical emission is in an average/low state during both TeV flares and this is
true also for the strictly simultaneous observations during the peak of the
flare in 2014. This can be explained by the harder-when-brighter behaviour
detected in X-rays, since the variability at lower energy has a lower
amplitude. On the other hand, the source is in high optical state in 2004 and
2006, when the TeV emission is very low. This suggests that at least two zones
or mechanisms are necessary to explain the flaring events in Mrk\,501.
In X-rays, the brightest event is recorded in the night of June 20-21, 2014
(MJD 56829), but does not show a correspondent maximum at VHE, as
might be expected. Given the rapid variability in both energy bands, this
could be explained by the non-simultaneity of the two observations. 
The TeV flux is roughly constant during this night and is $\sim$30\% higher than the one recorded
 on June 12, 2012. On this date, almost simultaneous X-ray observations show a $\sim$55\% lower X-ray flux.
Assuming an SSC emission model, a quadratic
correlation between the two energy bands is expected. Thus, the source should
be able to increase its $\gamma$-ray emission of almost a factor 4 in $\sim$\,90
minutes. This can easily be acieved thanks to a flux doubling timescale shorter
than 10 minutes \cite{nachi_icrc_proc_2015}.
A similar behaviour of high X-ray and low $\gamma$-ray emission is detected on
the night of June 22-23, 2014. In this case, though, the TeV flux roughly
doubles in the $\sim$\,2 hours of observations, suggesting a further increase
to match the X-ray flux state. The opposite behaviour is registered during the 
TeV flare on June 23-24, 2014 (MJD 56831). The X-ray emission is $\sim$\,20\%
lower than the maximum reached three nights before, while the $\gamma$-ray
flux is $\sim$\,4 times higher.
 
More details on the flux variability in 2014 can be found in \cite{nachi_icrc_proc_2015}.

\begin{figure}[tp]
  \centering
  \includegraphics[width=0.49\columnwidth]{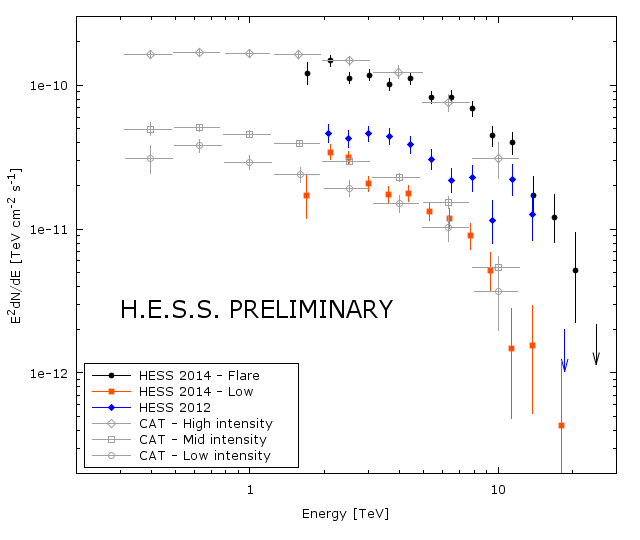}
  \includegraphics[width=0.49\columnwidth]{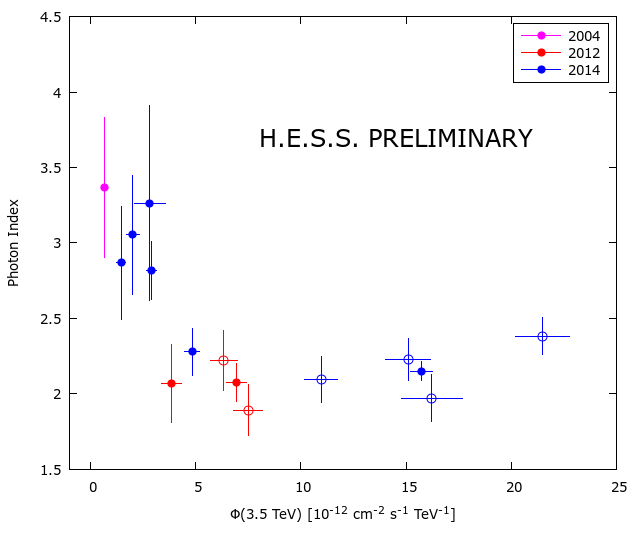}
  \caption{\textit{Left}: H.E.S.S. observed spectra for the 2014 and 2012 flare and low
    states. The 1997 CAT spectra \cite{CAT_1997_1999A&A...350...17D} are also shown for
    comparison. \textit{Right}: Intrinsic photon indices from EBL absorbed PL
    fits as function of flux. Values are given for single nights (full circles) and for single
    runs (open circles) during the flares. By high fluxes, the spectra are clearly hard.}
  \label{fig:gamma_vs_flux}
\end{figure}

\textbf{TeV spectra.} The very different flux states and high significance of
the detection enable a temporally resolved spectral analysis down to very
short timescales. Flux dependent spectral variability is described in the
literature (e.g.,
\cite{CAT_1997_1999A&A...350...17D,MAGIC_2005_2007ApJ...669..862A}), although
it has been determined mostly at energies below 2\,TeV. Analyses of several
datasets have been performed (total dataset, single nights and single runs
during the flares) and the 2012 and 2014 spectra are shown in
Fig.\,\ref{fig:gamma_vs_flux} (\textit{left}). Power law (PL) spectral fits provide in general poor
results and are acceptable only for low-flux nights and for some run-wise
analysis. This is expected from the curved spectra described in the literature
(e.g. \cite{CAT_1997_1999A&A...350...17D,HEGRA1997_spectrum_1999A&A...349...11A,MAGIC_2005_2007ApJ...669..862A})
and from the consideration that the intrinsic emitted spectrum must have been
absorbed through the interaction with the extragalactic background light
(EBL). Mrk\,501 spectrum extends up to $\sim$20\,TeV. Despite being a
  close source, a significant optical depth for EBL absorption $\tau\gtrsim$1 is expected.
Hence, although the curved power law (CPL) and exponential cut-off power law
(ECPL) spectral shapes fit the data satisfactorily, they are not a good
description of the intrinsic spectra. For this reason, in order to determine
the source intrinsic spectra in the various flux states, fits including EBL
absorption have been performed. The model of
\cite{Franceschini_2008A&A...487..837F} has been used. The spectra are best
described by an EBL-absorbed PL. The photon indices are $2.3\pm0.1$ and
$2.15\pm0.06$ for the 2012 and 2014 flaring states, and $2.7\pm0.1$ for the
2014 low state. The respective normalizations at 3.5\,TeV are $6.7\pm0.3$,
$15.7\pm0.5$ and $2.8\pm0.1$ cm$^{-2}$s$^{-1}$TeV$^{-1}$. Fits involving more
complex spectral models are generally not significantly better anymore.

The inspection of the residuals of night and run fits shows, however, a
possible trend towards a spectral softening at the highest energies,
especially when lower fluxes are considered. The impossibility to detect an
actual intrinsic curvature can be due to the too narrow energy range of about
one decade covered by the spectra.
On the other hand, the fact that this trend is mostly visible at lower fluxes
points towards a harder-when-brighter behaviour also at these high
energies. Indeed, the plot of the changes of the intrinsic photon
indices as function of flux (Fig.\,\ref{fig:gamma_vs_flux}, \textit{right}) shows that the
spectra are clearly hard during the flares. This
can be explained as being in the vicinity of the maximum of the IC peak, which
is known to migrate towards higher energies during flares and periods of high
activity (e.g. \cite{CAT_1997_1999A&A...350...17D,MAGIC_2005_2007ApJ...669..862A}). 
This is the first time that such behaviour is reported for energies
extending significantly beyond 10\,TeV, up to 20\,TeV and for such short time
intervals.
The hard PL intrinsic spectrum during the 2014 flare also indicates that 
there is no sign of being in the Klein-Nishina regime even at these high energies.

\textbf{EBL determination.} 
It is possible that the optical depth $\tau$
derived from \cite{Franceschini_2008A&A...487..837F} for the 2-20\,TeV energy
range at the redshift of Mrk\,501 ($z\simeq$\,0.034) is not
  precise. In order to test the model, the procedure described in
\cite{ebl_paper_2013A&A...550A...4H} was applied.
The sum of the LogLikelihood ratio test\footnote{LLRT = $-2 \cdot
  log[\mathcal{L}(H_0)/\mathcal{L}(H_1)]$, where $H_0$ is the null hypothesis,
  i.e. $\alpha=0$, and $H_1$ the hypothesis to be tested, $\alpha>0$. The
  significance is then calculated as $\sqrt{LLRT}$.} (LLRT) profiles peaks at the
normalization factor $\alpha=0.93^{+0.15}_{-0.14}$, 
with a value of $\sim$83, which translate in a significance 
of 9.1\,$\sigma$. The null hypothesis of a non-existent EBL is hence excluded 
at more than 9\,$\sigma$ level. The derived $\alpha$ is compatible with the model 
of \cite{Franceschini_2008A&A...487..837F}.
Taking into account the contributions of the two flares only, $\alpha = 0.89^{+0.16}_{-0.14}$,
which is not compatible with the one calculated in \cite{ebl_paper_2013A&A...550A...4H}
($\alpha=1.27^{+0.18}_{-0.15}$). This can be explained by the fact that,
altough several sources have been used in
\cite{ebl_paper_2013A&A...550A...4H}, the major contribution is given by
PKS\,2155-304 alone. Differences could therefore be explained as dependencies
on the redshift ($z_{Mrk\,501}\simeq0.034$ vs. $z_{PKS\,2155-304}=0.116$) and
on the significantly different energy range covered ($\sim$\,2\,-\,20\,TeV
vs. $\sim$\,0.15\,-\,7\,TeV). A further comparison with the low-redshift
subset composed of Mrk\,421 and PKS\,2005-489 only (<$z$>\,=\,0.051) is not
very meaningful because of the large errors on the normalization
($\alpha_{low-z} = 1.6^{+0.5}_{-1.1}$, \cite{ebl_paper_2013A&A...550A...4H})
and because the various subsets yield results
with LLRT values between -2 and 2, hence not significant. On the other hand, the
present result is perfectly compatible with the one derived in
\cite{ebl_paper_2013A&A...550A...4H} for the high energy dataset
$\alpha_{high-energy} = 1.05^{+0.32}_{-0.28}$, which covers the range 0.95\,-\,14\,TeV.

%%%%%%%%%%%%%%%%%%%%%%%%%%%%%

\textbf{LIV studies and QG scale limits.} The 2014 flare data were exploited
to search for Lorentz Invariance Violation (LIV) effects leading to limits on
the Quantum Gravity (QG) scale. In spite of its low redshift ($z\sim$\,0.034),
Mrk\,501 is a very promising source for LIV studies due to the hard energy
spectrum. 
The study of energy dependent time-delays within a deterministic scenario and
with a Likelihood method was performed following
\cite{liv_pks2155_2011APh....34..738H}. 
It consists of the construction of a low-energy
template light-curve (2\,-\,4.5\,TeV) injected in the likelihood fit
procedure with photons of energy above 4.5\,TeV ($\sim$\,500 photons with a
negligible background contribution below 0.5\%) and studied 
with dedicated simulations. Two cases were considered: a
linear and a quadratic dependence of the speed of light on the photon
energy. As no significant time delay was found (more than
1\,$\sigma$ from zero), limits on the QG scale were computed.
The one-sided 95\% CL QG limits including systematic uncertainties for the
sub-luminal (supra-luminal) case are 8.5$\times$10$^{17}$\,GeV
(6.4$\times$10$^{17}$\,GeV) for the linear term and 1.15$\times$10$^{11}$\,GeV
(1.0$\times$10$^{11}$\,GeV) for the quadratic term. The main uncertainties come
from the template light-curve binned fit, energy calibration, knowledge of the
acceptance corrections and of the smearing factors in energy. The total
contribution of the systematic effects to the error calculation was
conservatively estimated to be below 100\% of the statistical one. Another
type of systematics may come from the energy-dependent source effects which
may lead to energy-dependent time lags, not considered in this study. 
As first results, these are one of best constraints on the linear term derived
from AGN observations (comparable with those obtained with the PKS\,2155-304
2006 flare \cite{liv_pks2155_2011APh....34..738H}) and the best values on the
quadratic term obtained until now with GRBs \cite{liv_grb_2013PhRvD..87l2001V}
and AGNs \cite{liv_pks2155_2011APh....34..738H} data. This is mainly due to the
exceptional strength of the Mrk\,501 2014 flare leading to a high number of
photons in the energy range of 10\,-\,20\,TeV.

\section{Conclusion}

H.E.S.S. observations of Mrk\,501 between 2004 and 2014 have been presented in 
a MWL context. Rapid variability in the 2\,-\,20\,TeV energy range during flaring 
states in 2012 and 2014 has been reported. No direct relation with 
the emission at lower energies has been found. While the different behaviour
at X-ray energies can be in principle explained by the time offset between
the observations (this would imply the capability of the source to increase its
$\gamma$-ray flux of a factor $\sim$four in $\sim$90 minutes), the optical emission
clearly does not correlate with the VHE flux. It is in low state in the
strictly simultaneous observations during the 2014 flare, while it is
significantly higher during the TeV low states in 2004 and 2006. This suggests
the presence of at least two zones or mechanisms able to produce flaring
events in this source. The TeV spectra are hard during the periods of high
flux, and softer otherwise. In particular, the hard simple PL shape extending
above 20\,TeV during the 2014 flare peak indicates the absence of
Klein-Nishina suppression at these high energies. Studies on the EBL intensity
exclude the hypothesis of a non-existent EBL at the 9$\sigma$ level. The
normalization factors derived with all datasets ($\alpha =
0.93^{+0.15}_{-0.14}$) and with the flare datasets only ($\alpha =
0.89^{+0.16}_{-0.14}$) are consistent with the model of
\cite{Franceschini_2008A&A...487..837F}. The quadratic limits on the QM scale
(1.15$\times$10$^{11}$\,GeV) are the best ones derived from GRB and AGN observations.

\section{Acknowledgments}
The support of the Namibian authorities and of the University of Namibia in
facilitating the construction and operation of H.E.S.S. is gratefully
acknowledged, as is the support by the German Ministry for Education and
Research (BMBF), the Max Planck Society, the German Research Foundation (DFG),
the French Ministry for Research, the CNRS-IN2P3 and the Astroparticle
Interdisciplinary Programme of the CNRS, the U.K. Science and Technology
Facilities Council (STFC), the IPNP of the Charles University, the Czech
Science Foundation, the Polish Ministry of Science and Higher Education, the
South African Department of Science and Technology and National Research
Foundation, and by the University of Namibia. We appreciate the excellent work
of the technical support staff in Berlin, Durham, Hamburg, Heidelberg,
Palaiseau, Paris, Saclay, and in Namibia in the construction and operation of
the equipment. 
N.C. acknowledges support from Alexander von Humboldt foundation.
O.M.K acknowledges financial support of the project FR/639/6-320/12 by the
Shota Rustaveli National Science Foundation under contract 31/77.
This research has made use of the lightcurves provided by the University of
California, San Diego Center for Astrophysics and Space Sciences, X-ray Group
(R.E. Rothschild, A.G. Markowitz, E.S. Rivers, and B.A. McKim), obtained at
\url{http://cass.ucsd.edu/~rxteagn/}.

\bibliographystyle{JHEP}
\bibliography{mrk_reduced}

\end{document}